\documentclass[energies,article,accept,pdftex,moreauthors]{Definitions/mdpi} 
\firstpage{1} 
\pubvolume{18}
\issuenum{18}
\articlenumber{3813}
\pubyear{2025}
\copyrightyear{2025}
\datereceived{20 March 2025} 
\daterevised{18 June 2025} 
\dateaccepted{21 June 2025} 
\datepublished{17 July 2025} 
\hreflink{https://doi.org/10.3390/en18143813} 
\usepackage{algorithm}
\usepackage{algorithmic}

\usepackage{optidef}
\Title{Incorporating a Deep Neural Network into Moving Horizon Estimation for Embedded Thermal Torque Derating of an Electric~Machine}

\TitleCitation{Incorporating a Deep Neural Network into Moving Horizon Estimation for Embedded Thermal Torque Derating of an Electric Machine}

\Author{Alexander Winkler 
 $^{1}$\orcidA{}, Pranav Shah $^{1}$\orcidB{}, Katrin Baumgärtner $^{2}$\orcidC{}, 
Vasu Sharma $^{1}$\orcidF{}, David Gordon $^{3}$\orcidD{} and~Jakob~Andert~$^{1,}$*\orcidE{} }

\AuthorNames{Alexander Winkler, Pranav Shah, Katrin Baumgärtner, Vasu Sharma, David Gordon and Jakob Andert}

\AuthorCitation{Winkler, A.; Shah, P.; Baumgärtner, K.; Sharma, V.; Gordon, D.; Andert, J.}

\address{%
$^{1}$ \quad Teaching and Research Area Mechatronics in Mobile Propulsion, RWTH Aachen University; Forckenbeckstr.~4, 52074 Aachen, Germany; alexander.winkler@rwth-aachen.de~(A.W.); pranav.shah@rwth-aachen.de~(P.S.); vasu.sharma@rwth-aachen.de~(V.S.); andert@mmp.rwth-aachen.de~(J.A.) 
    \\
$^{2}$ \quad IMTEK---Department Of Microsystems, University of Freiburg, Georges-Köhler-Allee 103, \mbox{79108 Freiburg im Breisgau,} Germany; katrin.baumgaertner@imtek.uni-freiburg.de
 \\
$^{3}$ \quad Donadeo Innovation Centre 
 for Engineering, Department of Mechanical Engineering, University of Alberta, 10th Floor, Edmonton, AB T6G 1H9, Canada; dgordon@ualberta.ca~ 
 }

\corres{Correspondence: andert@mmp.rwth-aachen.de}

\abstract{This study presents a novel state estimation approach integrating Deep Neural Networks (DNNs) into Moving Horizon Estimation (MHE). This is a shift from using traditional physics-based models within MHE towards data-driven techniques.
Specifically, a Long Short-Term Memory (LSTM)-based DNN is trained using synthetic data derived from a high-fidelity thermal model of a Permanent Magnet Synchronous Machine (PMSM), applied within a thermal derating torque control strategy for battery electric vehicles.
The trained DNN is directly embedded within an MHE formulation, forming a discrete-time nonlinear optimal control problem (OCP) solved via the \texttt{acados
} optimization framework.
Model-in-the-Loop simulations demonstrate accurate temperature estimation even under noisy sensor conditions and simulated sensor failures.
Real-time implementation on embedded hardware confirms practical feasibility, achieving computational performance exceeding real-time requirements threefold.
By integrating the learned LSTM-based dynamics directly into MHE, this work achieves state estimation accuracy, robustness, and adaptability while reducing modeling efforts and complexity.
Overall, the results highlight the effectiveness of combining model-based and data-driven methods in safety-critical automotive control systems.}

\begin{document}

\section{Introduction}

State estimation has been a critical area of research for several decades, playing a fundamental role in various engineering disciplines by providing robust and reliable feedback for control systems~\cite{Simon2006}. 
The mathematical foundations of state estimation can be traced back to Carl Gauss in the early 1800s, with further advancements in the 20th century, including Maximum Likelihood Estimation~\cite{Aldrich97} and Linear Minimum Mean-Square Estimation~\cite{Janacek75}.
However, the applicability of these early formulations to real-time systems was limited by available computational resources.
As the focus of both academia and industry moved to non-linear systems, extensions such as the Extended Kalman Filter (EKF) and Unscented Kalman Filter (UKF) were developed to address non-linearities in state estimation~\cite{ribeiro04,  Julier97}. 
More recently, research has focused on state estimation techniques that can incorporate constraints on states and parameters, like Moving Horizon Estimation (MHE)~\cite{Rawlings2001, Vandersteen2013, Bae03072017, Baumgaertner2019, Baumgaertner2021, Girrbach2021}.

\begin{figure}[H]
	\centering	
    \includegraphics[height=0.3\textheight]{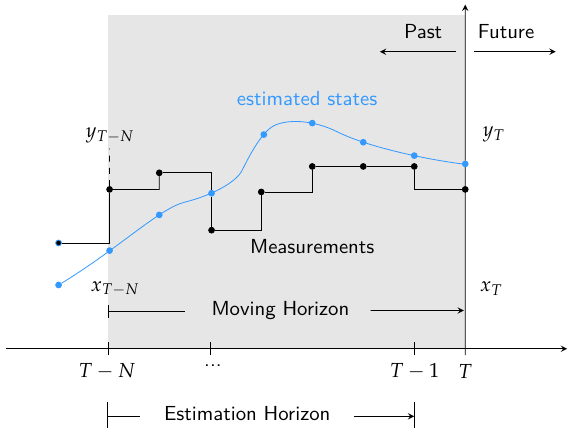}
	\caption[{Schematic}  of moving horizon estimation problem.]{Schematic 
 of MHE problem.}
	\label{fig:MHE_concept_intro}
\end{figure}

The main idea of MHE is to compute a state estimate $x$ by solving an optimization problem over a finite time horizon.
At each sample time, as soon as a measurement $y$ is made available, the horizon is shifted ahead in time in a receding horizon fashion and a new estimate is calculated.
Figure~\ref{fig:MHE_concept_intro} illustrates the MHE's operating principle using the receding horizon.

Compared to Kalman-based approaches, MHE offers improved handling of constraints, disturbances, and system non-linearities, making it a powerful tool for advanced estimation tasks~\cite{Rawlings2021}.

A critical aspect of modern state estimation is data assimilation, which integrates sensor measurements with dynamic process models to improve the accuracy of the estimation~\cite{Asch16}. 
Traditional approaches rely on physics-based models, which require detailed system knowledge and intricate mathematical formulations.
However, as systems become increasingly complex, the development of precise mathematical models becomes a challenge.
In addition, accurate models might be computationally too expensive in particular for real-time applications~\cite{brunton19}. 
These models often involve high-dimensional equations that demand significant computational resources, making them less practical for embedded control systems with stringent latency constraints. 
To address these challenges, data-driven approaches such as system identification, machine learning, and deep learning have gained prominence~\cite{Goodfellow16, hewing2020learning, salzmann2024learning, lahr2024l4acados}.

Deep learning architectures, inspired by biological neural networks, progressively build representations of input data while minimizing the need for manual feature engineering~\cite{lecun1998, bishop}.
The layers in a Deep Neural Network (DNN) consist of neurons that integrate weighted inputs via activation functions.
This function introduces non-linearity that is essential for learning complex tasks.
While the basic calculations within the neurons remain consistent, the arrangement of the neurons plays a crucial role in how the information is processed, giving rise to various neural network architectures.
Among these, Recurrent Neural Networks (RNNs) are particularly suited for dynamic systems due to their ability to retain temporal information~\cite{Sarker21}.
Long Short-Term Memory (LSTM) networks, a variant of RNNs, are especially effective for time-series modeling, as they mitigate vanishing gradient issues and capture long-term dependencies~\cite{hochreiter1997}.

The rapid advancement of DNNs has enabled the modeling of complex non-linear relationships without explicit system equations, making them invaluable for state estimation in systems where deriving accurate mathematical models is challenging.
DNN-based models provide a scalable alternative by learning system behavior directly from data, eliminating the need for complex physics-based formulations and enabling efficient real-time deployment.
Several studies have successfully integrated DNNs into 
{other state estimation frameworks, with applications in pseudo-measurement generation, parameterized EKF formulations, and RNN-based non-linear system identification~\cite{Bragantini21, Suykens95, pan00}.}

{Since DNNs can learn complex dependencies without prior advanced feature engineering, problems like MHE stand to benefit directly from data-driven predictors.
Incorporating DNNs into the MHE framework has broadly be categorized under three categories, i.e., improving the model quality, adapting the optimization cost, or leveraging the universal approximation capabilities of DNNs to replace the MHE itself \cite{mobeen2025neural}.
Promising results from previous studies have successfully applied this approach to different applications~\cite{Song.2023,Mostafavi.18112022,Chen.2021, Alessandri.129200812112008}.
Ref.
~\cite{Chen.2021} developed a two-step optimization strategy where an offline learnt autoregressive LSTM model is used to estimate the state of charge of lithium-ion batteries, which is later improved using MHE through online optimization. Ref.~\cite{Mostafavi.18112022} incorporated the MHE estimates obtained using a feed-forward network for unknown heating, ventilation, and air conditioning (HVAC) dynamics into an model predictive control (MPC). 
Despite the common optimization-based foundations of MPC and MHE, the direct integration of learned dynamic models, particularly using LSTM and DNN architectures, has predominantly been explored within control applications.
Notable examples include MPC for combustion engines~\cite{NOROUZI2022CEP} and rapid development toolchains for low-temperature combustion processes~\cite{Gordon2024Intro}.
However, directly embedding a learned dynamics model in the form of a DNN into MHE remains unexplored.}

{In previous work \cite{WINKLER20238254}, an MPC was introduced for thermal torque derating of Permanent Magnet Synchronous Machines (PMSMs) in Battery Electric Vehicles (BEVs).
PMSMs are widely used for automotive propulsion due to their high torque, power density, and efficiency.
However, these advantages come with substantial localized heat generation, particularly under high-load scenarios, increasing the risk of permanent magnet demagnetization and insulation damage.
To prevent component failures and avoid cooling system over-dimensioning, thermal derating strategies limit torque output as critical temperatures are approached~\cite{Engelhardt.2017}.
Conventional, rule-based derating methods, such as linear mappings, are straightforward but neglect the thermal dynamics, resulting in overly conservative operation~\cite{Etzold.2019b, Wallscheid.2017}.
Effective thermal derating thus requires anticipation of future driving conditions and accurate modeling of the machine's thermal response to prevent dangerous temperature overshoots.
Previous research \cite{WINKLER20238254} focused on optimizing torque commands with a DNN-based MPC, ensuring adherence to PMSM temperature constraints within a Model-in-the-Loop (MiL) simulation, guided by a reference velocity profile derived from a Nürburgring Nordschleife lap.}

{Furthermore, these control methods are highly dependent on accurate sensor feedback data.
Therefore, accurate state estimation is crucial.
Building upon the prior work, this study proposes a DNN-based MHE to provide reliable feedback, enhance control accuracy, and ensure robust protection of the machine.
The novelty of this study lies in using a learned LSTM-based dynamics model within MHE, replacing traditional white- or grey-box approaches as the MHE dynamics model.
This integration aims to enhance state estimation accuracy, adaptability, and robustness, while simultaneously reducing modeling complexity and computational effort.}

\begin{figure}[H]
    	\includegraphics[width=1\textwidth]{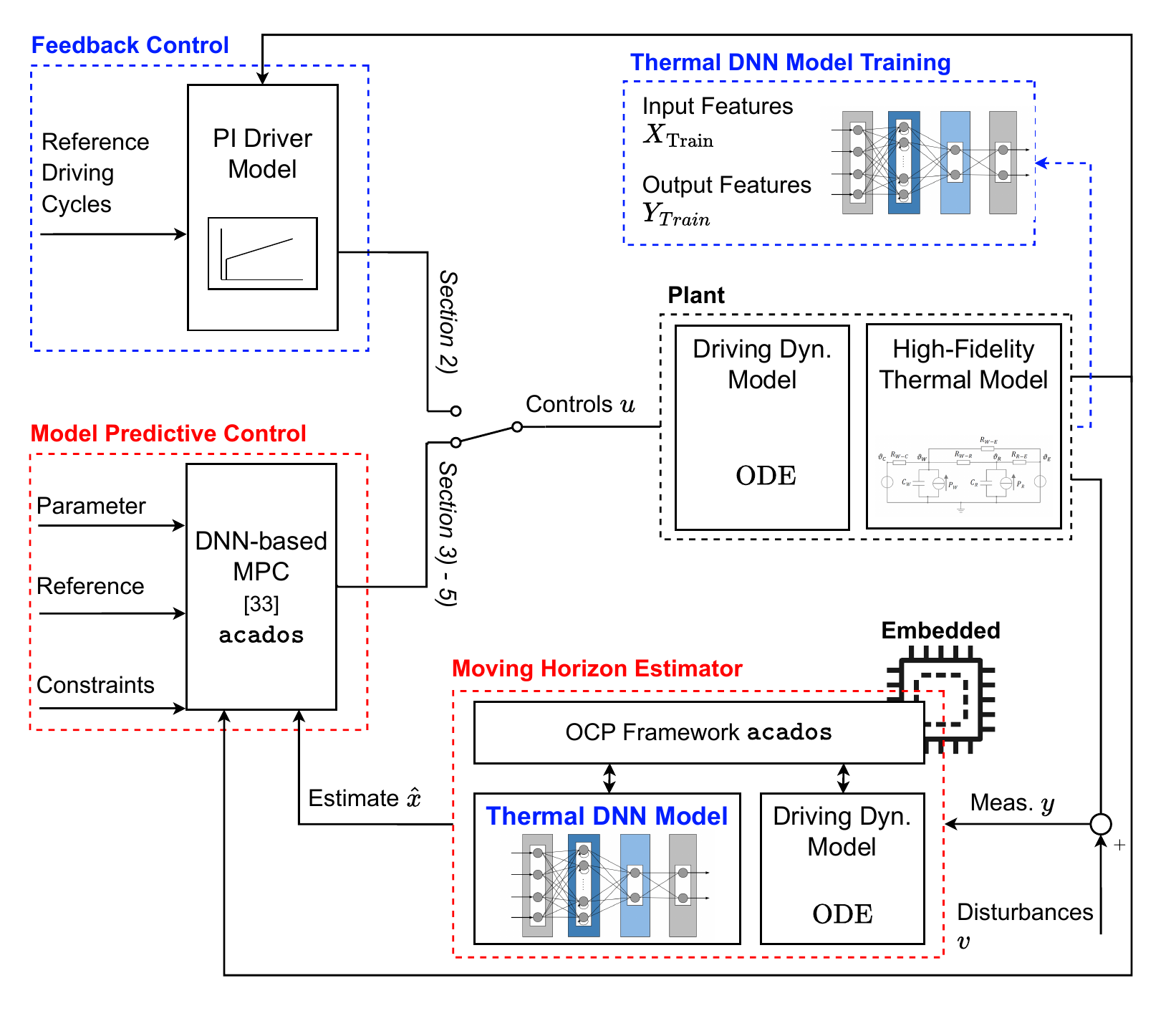}
	\caption[Graphical abstract.]{{Graphical abstract summarizing the work, including synthetic data generation. DNN-based MPC as presented in Ref.~\cite{WINKLER20238254}.}}
    \label{fig:GA}
\end{figure}

{To provide a high-level overview of the approach and methodology, the graphical abstract in Figure~\ref{fig:GA} illustrates the key components of the proposed MHE framework.
It shows the workflow of synthetic data generation using a high-fidelity, experimentally validated Lumped Parameter Thermal Network (LPTN) model, followed by DNN training for thermal state prediction of the PMSM.
The trained DNN model is then integrated into the MHE within the MiL simulation setup, where the estimator processes noisy sensor measurements to reconstruct accurate system states for the MPC.
The MHE, implemented using the \texttt{acados} optimal control framework for embedded compatibility, is then deployed on an embedded platform to assess its real-time capability.}

{The central objective and nature of this research is to demonstrate the feasibility of a DNN-based MHE framework.
Rather than conducting quantitative comparisons of vehicle performance under MPC control, this study provides a qualitative assessment of the DNN-based MHE’s effectiveness in estimating temperature states and its robustness against injected sensor faults.}

 By demonstrating the feasibility of integrating DNN-based MHE within an MPC-controlled system, this research paves the way for next-generation data-driven state estimation strategies.
 The fusion of deep learning with real-time optimal control unlocks new potential for intelligent, adaptive, and computationally efficient estimation frameworks, bridging the gap between model-based and purely data-driven approaches.

The paper is divided into the DNN modeling (Section~\ref{sec:DNN}), the MHE development (Section~\ref{sec:MHE}), and the subsequent simulative and embedded integration for validation (Section~\ref{sec:SimResults} and Section~\ref{sec:EmbeddedIntegration}, respectively) sections as shown schematically in Figure~\ref{fig:GA}.

This research makes the following key contributions:
\begin{enumerate}
    \item {Integration of DNN with MHE: 
} {This work represents one of the first attempts to directly integrate a DNN-based model within MHE, offering a novel framework for state estimation.
    Instead of relying on DNNs externally to provide Supplementary Information, the MHE leverages the DNN’s learned dynamics directly within the optimization.} 
    \item {Deployment and validation on embedded systems: 
} The feasibility of implementing a DNN-based MHE framework in real time is demonstrated.
    This research is also one of the first to test and validate the deployment of DNN-based state estimation on real-time hardware using the \texttt{acados} optimal control framework~\cite{Verschueren2021}.
\end{enumerate}

\section{Materials and Methods}
\label{sec:MaterialMethods}
{ 
This section describes the development of the DNN model for thermal state prediction and its subsequent integration into the MHE framework.
First, the LSTM network architecture, synthetic data generation, and training procedure are detailed.
Then, the trained DNN model is incorporated into the estimator by reformulating both driving dynamics and thermal dynamics into a discrete-time optimization structure compatible with \texttt{acados}.
}

\subsection{Deep Neural Network Modeling}
\label{sec:DNN}
\subsubsection{Long Short-Term Memory Network}
\label{sec:DNN_LSTM}

LSTMs, one of the most popular variants of RNN, have the ability to maintain a form of memory, enabling them to influence current inputs and outputs based on past sequence information~\cite{Sarker21}.
To regulate the flow of data, LSTMs introduce memory cells and gate units~\cite{hochreiter1997}.
The core computational unit of the LSTM network is called the memory cell (or simply ``cell''), and these networks are primarily designed for sequence modeling while addressing the vanishing gradient problem~\cite{Sarker21, Brownlee17}.

Each LSTM cell consists of three gates: the forget gate, the input gate, and the output gate that regulate information flow using sigmoid ($\sigma$) and hyperbolic tangent ($\tanh$) activation functions to maintain stable outputs~\cite{Sarker21}. The following equations mathematically define the operations within an LSTM cell, where each gate contributes to memory updates and output generation: 
\begin{equation}\label{eq:LSTMeq}
\begin{split}
    i_k &= \sigma\left(W_{\mathrm{u,i}}^T u_k + W_{\mathrm{h,i}}^T h_{k-1} + b_{\mathrm{i}}\right)~, \\
    f_k &= \sigma\left(W_{\mathrm{u,f}}^T u_k + W_{\mathrm{h,f}}^T h_{k-1} + b_{\mathrm{f}}\right)~, \\
    g_k &= \tanh\left(W_{\mathrm{u,g}}^T u_k + W_{\mathrm{h,g}}^T h_{k-1} + b_{\mathrm{g}}\right)~, \\
    o_k &= \sigma\left(W_{\mathrm{u,o}}^T u_k + W_{\mathrm{h,o}}^T h_{k-1} + b_{\mathrm{o}}\right)~, \\
    c_k &= f_k \odot c_{k-1} + i_k \odot g_k~, \\
    h_k &= o_k \odot \tanh\left(c_k\right)~,
\end{split}
\end{equation}
where $W_{\mathrm{u,(f,g,i,o)}}$ are the weighting matrices applied to the input vector $u_k$. $W_{\mathrm{h,(f,g,i,o)}}$ are weight matrices of the previous hidden state~$h_{k-1}$.
In this equation, \(\odot\), is an element-wise multiplication and \(b_{(f,g,i,o)}\) are the biases.
$i_{k}$ is the input gate, $f_{k}$ is the forget gate, $g_{k}$ is the cell candidate, $o_{k}$ is the output gate, $c_{k}$ is the cell state, and $h_{k}$ is the hidden state. 
Two activation functions are used in Equation~(\ref{eq:LSTMeq}) which are given as

\begin{enumerate}
    \item $\tanh(z)$ hyperbolic tangent activation function:
    \begin{equation}\label{eq:tanh}
        \tanh(z) = \frac{e^{2z}-1}{e^{2z}+1}~,
    \end{equation}
    \item $\sigma(z)$ sigmoid activation function:    \begin{equation}\label{eq:sig}
        \sigma(z) = \frac{1}{1 + e^{-z}}~.
    \end{equation}
\end{enumerate}

\begin{figure}[H]
\includegraphics[width=12 cm]{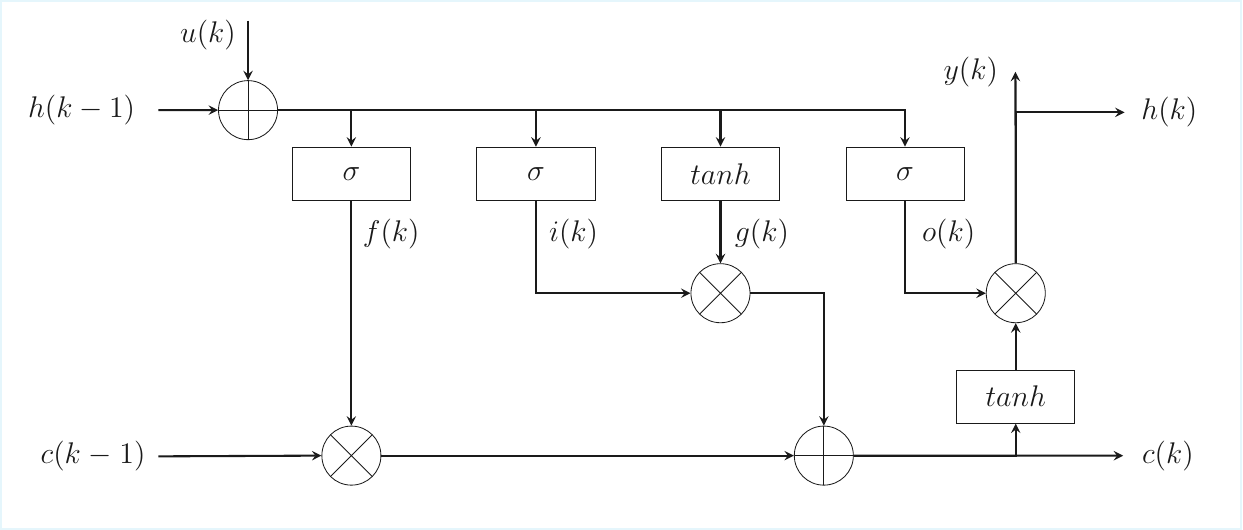}
    \caption{LSTM unit. 
  $\oplus$ Element-wise addition, $\otimes$ element-wise multiplication, $u$: input, $y$: output, $c$: cell state, $h$: hidden state, $\sigma$: sigmoid function, $\mathit{tanh}$: hyperbolic tangent function.}
	\label{fig:LSTM_graph}
\end{figure}

These activation functions are used to introduce non-linearity into the otherwise linear layers.
Figure~\ref{fig:LSTM_graph} shows a visualization of the information flow depicted in Equation~(\ref{eq:LSTMeq}).

Due to the temporal dependencies in LSTMs, storing both the cell and hidden states across timesteps increases memory requirements.

\subsubsection{Experimental Setup and Data Generation}
\label{sec:DNN:DataGen}
The proposed state estimator, akin to the approach of previous works in~\cite{WINKLER20238254}  uses one-dimensional white-box driving dynamics and thermal black-box DNN dynamics to predict the thermal states of a PMSM and generate accurate and robust estimates~\cite{WINKLER20238254,  WINKLER2021359}.
The DNN, integrated with the state estimator, predicts the temperature gradients of the PMSM to calculate accurate thermal states of the system.
For simplification, the PMSM is represented by two thermal masses (\(\theta_{\mathrm{w}}, \theta_{\mathrm{r}}\)) corresponding to measurable real-world temperatures at the test-bench of the windings and the rotor, respectively.
When combined with the MPC framework, the resulting MiL simulation enables safe thermal derating of a BEV by effectively handling noisy PMSM temperature measurements while ensuring compliance with thermal constraints.

The performance and accuracy of the DNN models heavily depend on the quality and quantity of training data.
To efficiently generate the required data without relying on extensive test-bench experiments, a simulation framework is employed.
Within this framework, a proportional--integral (PI) driver controller computes the torque requests to the Electric Machine (EM) and friction brake based on the vehicle velocity $v$.
These include (\(T_{\mathrm{EM,acc}},\) \(T_{\mathrm{EM,brk}},\) \(T_{\mathrm{fric,brk}}\)), where the torque of the EM \( (T_{\mathrm{EM}}) \) is split into acceleration and braking components: $T_{\mathrm{EM}}=T_{\mathrm{EM,acc}}+T_{\mathrm{EM,brk}}$.
A one-dimensional longitudinal model utilizes these torque values to calculate vehicle velocity $v_{\mathrm{veh}}$ utilizing the ordinary differential equation (ODE):
\begin{equation}
\label{eq:drivingDynamics}
    \begin{split}
        \dot{v}_{\mathrm{veh}} & = m_{\mathrm{veh}}^{-1}\cdot ( (T_{\mathrm{EM,acc}} + T_{\mathrm{EM,brk}} + T_{\mathrm{fric,brk}}) \cdot i_{\mathrm{diff}}\cdot r_{\mathrm{dyn}}^{-1} - m_{\mathrm{veh}}\cdot g\cdot \sin(\phi) \\ & - 0.5\cdot c_{\mathrm{d}}\cdot A_{\mathrm{c}}\cdot \rho\cdot v_{\mathrm{veh}}^{2} - m_{\mathrm{veh}}\cdot g\cdot \cos(\phi)\cdot c_{\mathrm{r}})~,
    \end{split}
\end{equation}
where $m_{\mathrm{veh}} = {1160}$ kg is the vehicle mass (including the driver), $c_{\mathrm{d}} = 0.32$ is the drag coefficient, $c_{\mathrm{r}} = 0.011$ is the rolling friction coefficient, $r_{\mathrm{dyn}} = {0.293}$ m is the dynamic tire radius, and $A_{\mathrm{c}} = {2.21}$~m$^2$ is the car's cross-sectional area.
The transmission ratio is given by $i_{\mathrm{diff}} = 9.3$, while the maximum vehicle speed is $v_{\mathrm{max}} = {130}$ km/h.
$\phi$ depicts an external parameter and input as the road inclination defined by the driving cycle, while $\rho$ and $g$ are the air density and gravitational acceleration, respectively.
In this study, a typical minicar BEV is used, and its parameters have been validated in previous works~\cite{WINKLER2021359}.

\begin{figure}[H]
	\centering
	\includegraphics[width=0.95 \textwidth]{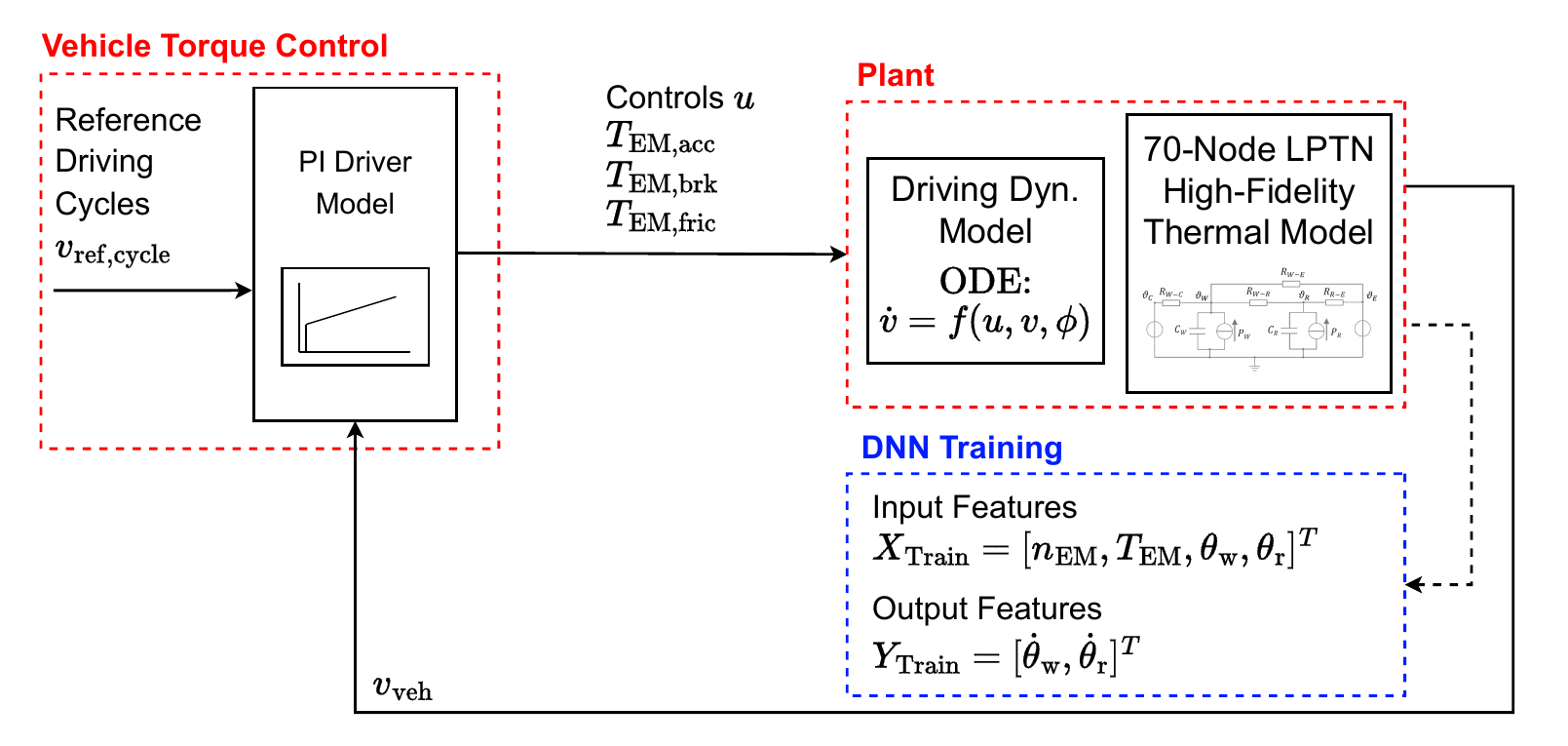}
	\caption[Simulation and model setup for training data generation]{Simulation and model setup for {synthetic} training data generation.}
	\label{fig:ANN_train}
\end{figure}

The relation between the vehicle speed $v$ in Equation~(\ref{eq:drivingDynamics}) and the rotational speed of the electric machine \(n_{\mathrm{EM}}\) is as follows:
$n_{\mathrm{EM}}=(30 \cdot v_{\mathrm{veh}} \cdot i_{\mathrm{diff}}) / (\pi \cdot r_{\mathrm{dyn}})$.
Thus, the primary operating point of the EM, the torque \(T_{\mathrm{EM}}\), and the rotational speed \(n_{\mathrm{EM}}\) serve as the primary inputs to the high-fidelity LPTN model, determining the rotor and winding temperature (\(\theta_{\mathrm{w}}, \theta_{\mathrm{r}}\)) of the machine, with their derivatives \((\dot{\theta}_w, \dot{\theta}_r\)) being monitored as well. 
The 70-node high-fidelity LPTN model is provided by the PMSM's manufacturer~\textit{DENSO}, and its parameters have been fitted using extensive experimental test-bench data.
The full simulation model, implemented in \texttt{MATLAB/Simulink}, is depicted in Figure~\ref{fig:ANN_train} and operates at a sample rate of {100} Hz.

\begin{figure}[H]
    \includegraphics[width=0.9 \textwidth]{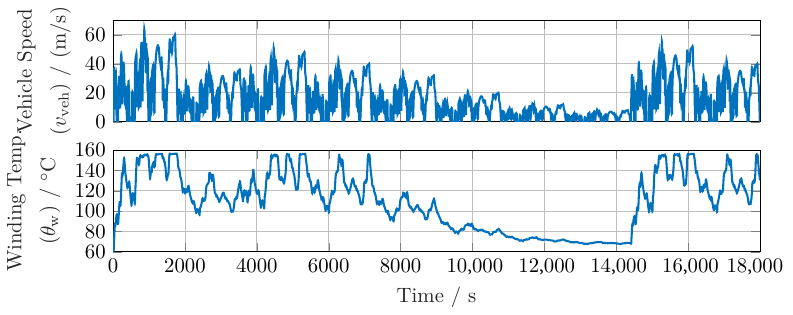}%
	\caption[Synthetic Data Generation]{{Synthetic 
 data for BEV vehicle speed and electric machine winding temperature, utilizing multiple drive cycles.}}
	\label{fig:DrivCycleTemp}
\end{figure}

A driving cycle serves as the primary predefined input for the data generation model, providing the reference velocity $v_\mathrm{ref}$ for the PI controller and the respective road inclination $\phi$.
To achieve higher and thus safety critical PMSM temperatures, the WLTP Class 3 driving cycle is customized, deliberately influencing the data distribution. Figure~\ref{fig:DrivCycleTemp} illustrates the velocity profile of the customized cycle and the corresponding temperature response, where only the winding temperature  $\theta_{\mathrm{w}}$ is shown, as the rotor temperature remains below critical levels and is therefore omitted.
The figure also highlights that a significant portion of the data lies within the machine's safety-critical temperature range, between {150} $^\circ$C and {160} $^\circ$C, where permanent damage can occur.

\subsubsection{Neural Network Training}
\label{sec:DNN:Train}
\begin{figure}[H]
	\includegraphics[width=0.95\textwidth]{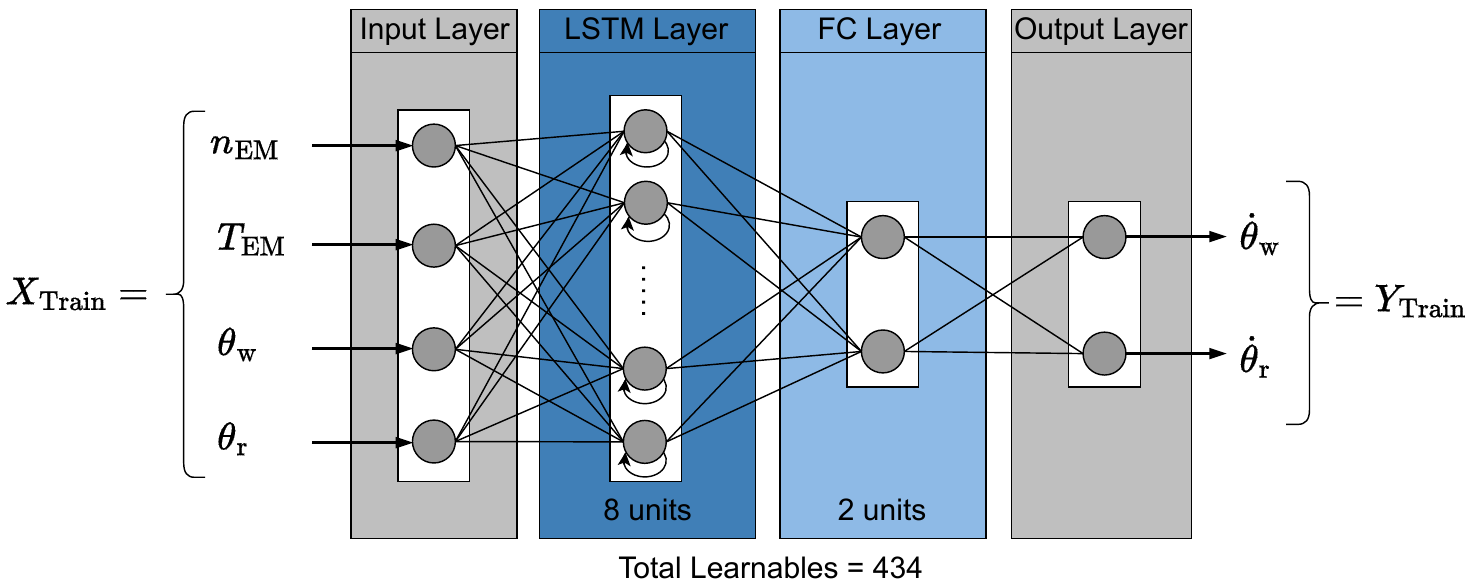}
  \caption[LSTM neural network architecture]{LSTM artificial neural network architecture using an LSTM and FC layer.}
  \label{fig: LSTM_arch}
\end{figure}

The network architecture comprises four layers: the input, LSTM, Fully Connected (FC), and output layer.
The LSTM layer consists of 8 LSTM cells that compute the temporal dependencies between the inputs.
Figure~\ref{fig: LSTM_arch} shows the architecture of the DNN used for training.

Drawing inspiration from LPTNs and the physical intuition behind heat transfer, this work models temperature evolution by predicting temperature change rates rather than absolute temperatures. 
Subsequently, the input features to the DNN are rotational speed of the EM \(n_{\mathrm{EM}}\), torque of the EM \(T_{\mathrm{EM}}\), winding temperature \(\theta_{\mathrm{w}}\), and rotor temperature \(\theta_{\mathrm{r}}\) of the EM.
The DNN's output features are the gradients \(\dot{\theta}_{\mathrm{w}}\) and \(\dot{\theta}_{\mathrm{r}}\), at timestep $k$.
This can be summarized as follows in Equation~(\ref{eq:DnnFeatures}):
\begin{align}
    \label{eq:DnnFeatures}
    X_{\mathrm{Train}}& = \begin{bmatrix}
        n_{\mathrm{EM}}(k) & T_{\mathrm{EM}}(k) & \theta_{\mathrm{w}}(k) & \theta_{\mathrm{r}}(k)
    \end{bmatrix}^{T} \quad \in\mathbb{R}^{4} ~, \notag \\ 
    Y_{\mathrm{Train}}& = \begin{bmatrix}
        \dot{\theta}_{\mathrm{w}}(k) & \dot{\theta}_{\mathrm{r}}(k)
    \end{bmatrix}^{T} \quad \in\mathbb{R}^{2} ~.
\end{align}

\begin{figure}[H]
    \includegraphics[width=0.95\textwidth]{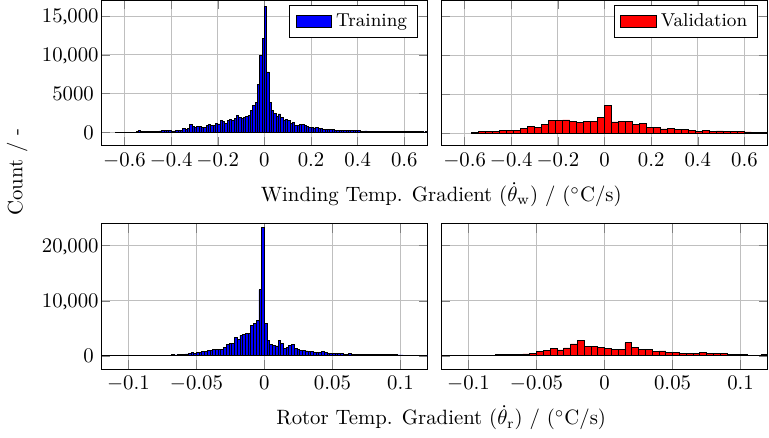}
    \caption[Data Distribution Rotor and Winding Temperature]{Data 
 distribution of network output gradient of electrical machine winding (upper plot) and rotor (lower plot) temperature. Training and validation dataset, $\mathcal{D}_{\text{train}}, \mathcal{D}_{\text{val}}$ (80:20 split). Total data points: 180,000.}
    \label{fig:TrwData}
\end{figure}

The depth of a DNN directly influences computational complexity, as each additional layer increases the number of learnable parameters and matrix operations.
To ensure real-time feasibility, particularly for integration with optimization techniques like MPC and MHE, the DNN's depth and the number of recurrent nodes are constrained.
The distribution of the {synthetic training data generated in Section~\ref{sec:DNN:DataGen}} is shown in Figure~\ref{fig:TrwData} for the two outputs $\dot{\theta}_{\mathrm{w}}$ and $\dot{\theta}_{\mathrm{r}}$. {The dataset, comprising 180,000 data points, is divided into a training set ($\mathcal{D}_{\text{train}}$) and a validation set ($\mathcal{D}_{\text{val}}$) in an 80:20 ratio.}

The training is conducted using the hyperparameters listed in Table~\ref{tab:hyperparam}, {with mean squared error (MSE) as the performance metric for supervised learning}.

\begin{table}[H]
 \caption{{Training hyperparameter settings.}}
    \label{tab:hyperparam}
    \setlength{\tabcolsep}{10.7mm}
    \renewcommand{\arraystretch}{1.2}
    \begin{tabular}{ll} 
        \toprule
        \textbf{Hyperparameter} &  \textbf{Value}\\
        \midrule
        Max epoch & 10,000 \\    
        {Performance metric} & {MSE} \\        
        Optimizer & Adam~\cite{kingma2014adam} \\
        Mini-batch size & 512\\
        Initial learning rate & 0.02\\
        Learn rate schedule & Piecewise drop by 25\% every 500 epochs\\
        L2 regularization & 0.1\\	
	Validation frequency & 10\\
        \bottomrule       
    \end{tabular}
\end{table}

The algorithm of the DNN training is depicted in Algorithm~\ref{alg:lstm_training_matlab}.

\begin{algorithm}[H]
	\caption{Pseudo-code: LSTM network training procedure using Adam optimizer.}
	\label{alg:lstm_training_matlab}
	\begin{algorithmic}[1]
		\REQUIRE Training dataset $\mathcal{D}_{\text{train}}$ (80\% of total dataset), validation dataset $\mathcal{D}_{\text{val}}$ (20\% of total dataset)
		\REQUIRE Network architecture as defined in Figure~\ref{fig: LSTM_arch}
        \REQUIRE $\mathcal{L} \leftarrow \text{MSE}$
		\ENSURE Trained network parameters $\theta^*$ (best validation loss)
		
		\STATE Initialize network parameters $\theta$ randomly
		\STATE Set optimizer, learning rate schedule, learning rate
		\STATE Set training hyperparameters: maxEpochs, miniBatchSize, gradientThreshold		
		\STATE Initialize record: $\mathcal{L}_{\text{val}}^{\text{best}} \leftarrow \infty$, $\theta^* \leftarrow \theta$
		
		\FOR{epoch $= 1$ to maxEpochs}
			\FOR{each mini-batch $\mathcal{B} \subset \mathcal{D}$ (sequential mini-batches)}
                \STATE Pad sequences in $\mathcal{B}$ to equal length
				\STATE Forward pass: compute predictions ${y}_{\text{pred}} = f_\theta(x_{\text{train}})$ for all sequences in $\mathcal{B}$
				\STATE Compute loss $\mathcal{L}({y}_{\text{pred}}, {y}_{\text{true}})$ over mini-batch
				\STATE Backpropagate gradients $\nabla_\theta \mathcal{L}$ (with gradient clipping at threshold 1)
				\STATE Update parameters $\theta$ using Adam optimizer
			\ENDFOR
			
			\IF{current iteration is multiple of 10}
				\STATE Evaluate validation loss $\mathcal{L}_{\text{val}}$ on $\mathcal{D}_{\text{val}}$
				\IF{$\mathcal{L}_{\text{val}} < \mathcal{L}_{\text{val}}^{\text{best}}$}
					\STATE Update best parameters: $\theta^* \leftarrow \theta$, $\mathcal{L}_{\text{val}}^{\text{best}} \leftarrow \mathcal{L}_{\text{val}}$
				\ENDIF
			\ENDIF
		\ENDFOR
		
		\STATE \textbf{return} best-performing network parameters $\theta^*$
	\end{algorithmic}
\end{algorithm}

\begin{figure}[H]
    \includegraphics[width=0.95\textwidth]{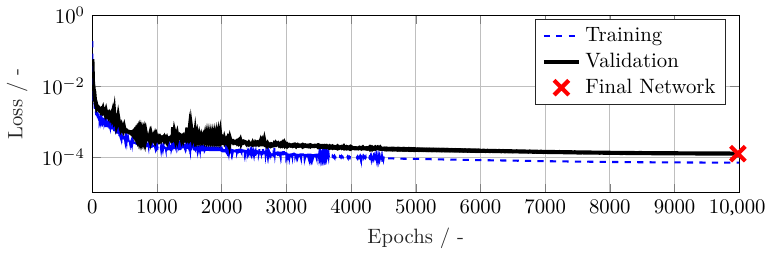}
    \caption[Loss-Epoch Plot Neural Network Training]{Loss-Epoch 
 plot for the neural network training. The final network is chosen according to the best training loss.}
    \label{fig:LossEpoch}
\end{figure}

\begin{figure}[H]
    \includegraphics[width=0.95\textwidth]{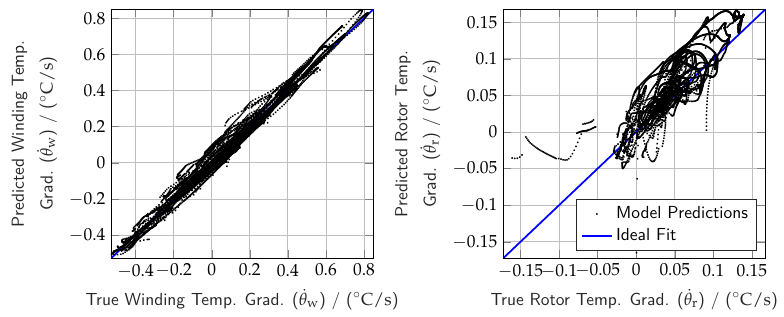}
    \caption[Predicted Network Outputs vs. Actual]{Predicted 
 neural network outputs vs. the actual ground truth data for both network outputs for unseen test dataset, gradients of winding and rotor temperature of the electric machine.}
    \label{fig:NetPredAct}
\end{figure}

The resulting loss-epoch plot of the training is shown in Figure~\ref{fig:LossEpoch}, while the prediction results of the final network on the unseen test dataset are depicted in Figure~\ref{fig:NetPredAct}.
The unseen test dataset consists of data for a simulated lap on the Nürburgring Nordschleife with its high power requirements, thus posing a major challenge to the electric drivetrain and its thermal constraints.
To solve the trade-off between the DNN's depth and thus the computational complexity and the DNN's prediction performance, empirical studies are performed, which are omitted here for brevity.
The results in Figure~\ref{fig:NetPredAct} show a good correspondence with the ground truth from the high-fidelity model, achieving an RMSE of {0.0373} $^\circ$C/s and {0.0282} $^\circ$C/s, and an NRMSE of {2.77}\% and {9.39}\% for $\dot{\theta}_{\mathrm{w}}$ and $\dot{\theta}_{\mathrm{r}}$, respectively. Further performance metrics are depicted in Table~\ref{tab:train_metrics}.

\begin{table}[H]
  \caption{{Deep neural network prediction performance metrics on unseen test dataset. MAE: Mean Absolute Error, RMSE: Root MSE, NRMSE: Normalized RMSE.}}
    \label{tab:train_metrics}
\setlength{\tabcolsep}{16mm}
    \renewcommand{\arraystretch}{1.2}
    \begin{tabular}{lll} 
        \toprule
        \textbf{Metric} &  \boldmath{$\dot{\theta}_{\mathrm{w}}$} &  \boldmath{$\dot{\theta}_{\mathrm{r}}$}\\
        \midrule
        MAE / ($^\circ$C/s) & 0.0288 & 0.0210 \\
        RMSE / ($^\circ$C/s) & 0.0373 & 0.0282 \\
        NRMSE / - & 2.77\% & 9.39\% \\
        \bottomrule
    \end{tabular}  
\end{table}
\vspace{-6pt}

%

The lower accuracy in predicting $\dot{\theta}_{\mathrm{r}}$ can be attributed to the complexity of the high-fidelity thermal model and the weaker influence of the selected features.
To be more precise, the network struggles with lower rotor temperatures due to a lack of training data in that range (see Figure~\ref{fig:TrwData}).
However, since the focus of data generation was on higher, critical temperatures, this limitation is acceptable, and the network is considered sufficiently accurate due to its superior performance on $\dot{\theta}_{\mathrm{w}}$.

\begin{figure}[H]
    \includegraphics[width=0.95\textwidth]{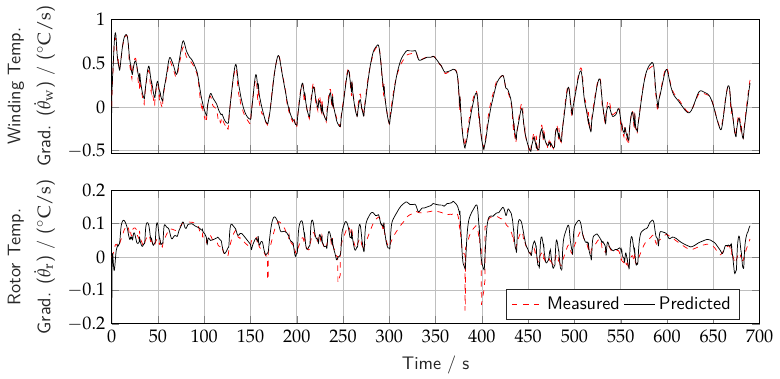}
    \caption[Predicted Network Outputs Time Domain]{Predicted 
 neural network outputs on the unseen test (Nürburgring Nordschleife) dataset over time domain: gradients of winding and rotor temperature of the electric machine.}
    \label{fig:NetPredTime}
\end{figure}

The presentation of the DNN predictions in the time-domain of the unseen test dataset of the Nürburgring Nordschleife in Figure~\ref{fig:NetPredTime} underlines the good performance of the chosen network.

Since the neural network predicts temperature gradients, explicit first-order Euler integration with an initial temperature of $\theta_{\mathrm{w,r},0} = {60}$ $^\circ$C for both the winding and the rotor is used to visualize the absolute temperatures.


\subsection{Estimator Formulation}
\label{sec:MHE}
MHE is employed to reconstruct system states by solving an optimization problem at each timestep. 
The optimization problem has a similar structure as an Optimal Control Problem (OCP) such that tailored solvers for OCP-structured problems can be used.
Following successful training of the DNN model, the next step involves formulating the estimator and implementing it within the OCP-structure of the \texttt{acados} framework.
Therefore, both the driving dynamics and the DNN-based thermal model are reformulated in a discrete-time, optimization-compatible form.

\subsubsection{MHE Problem Formulation}

The dynamics depicted here are based on a discrete-time, non-linear, time-invariant system~\cite{Rawlings17}:
\begin{align} \label{Eq. nl_discrete}
    x_{k+1} = \Tilde{f}( x_{k}, w_{k}), \qquad y_{k} = g( x_{k}) + v_{k} ~,
\end{align}
with state \(x \in \mathbb{X} \subseteq \mathbb{R}^n\), measurement \(y \in \mathbb{Y} \subseteq \mathbb{R}^{p}\), process disturbance \(w \in \mathbb{W} \subseteq \mathbb{R}^{g}\), measurement disturbance \(v \in \mathbb{V} \subseteq \mathbb{R}^p\). 
The non-linear function \(\Tilde{f}(x_k, w_k)\) describing the system dynamics and the measurement model \(g(x)\) are now developed, bringing the DNN-based thermal model and the one-dimensional driving dynamics model in a common, discrete form.

Firstly, the driving dynamics in Equation~(\ref{eq:drivingDynamics}) can be further summarized and discretized using explicit Euler integration of first order with integration interval $\delta k$ as

\begin{equation}
\label{eq:drivingDynamics_func_discr}    
        v_{\mathrm{veh},k+1} = v_{\mathrm{veh},k} + f_{\mathrm{DD}} ( v_{\mathrm{veh},k}, T_{\mathrm{EM,acc},k}, T_{\mathrm{EM,brk},k}, T_{\mathrm{fric,brk},k}, \phi_{k} ) \cdot \delta k ~,  
\end{equation}
with $f_{DD}$ summarizing the driving dynamics equation.

Equation~(\ref{eq:LSTMeq}) presents the equations for an LSTM unit. 
This formulation is transformed into a forward-propagating process model by representing operations as equations incorporating stored weights and biases.
The LSTM internal memory states---the hidden and cell states $(c, h)$---are included in the network inputs and outputs due to the unrolling of the recurrent layer.
Adding the FC layer of the DNN provides the full dynamics representation of the DNN: 
\begin{align} \label{Eq:DNN_FC_equ}
    \begin{bmatrix}
        \delta\theta_{\mathrm{w},k+1} & \delta\theta_{\mathrm{r},k+1} & h_{k+1} & c_{k+1}
    \end{bmatrix}^T &= f_{\mathrm{DNN}}(
    \theta_{\mathrm{w},k}, \theta_{\mathrm{r},k}, h_{k}, c_{k} ) ~,
\end{align}
where 
 function $f_{\mathrm{DNN}}$ thus summarizes the complete DNN dynamics.

Finally, the absolute temperature predictions for the next timestep using explicit first-order Euler integration are defined as
\begin{align}  \label{Eq:thermal_dyn_final}
    \begin{bmatrix} 
        \theta_{{\mathrm{w},k+1}} & \theta_{{\mathrm{r},k+1}}
    \end{bmatrix}^{T} = 
    \begin{bmatrix}
        \theta_{\mathrm{w},k} & \theta_{\mathrm{r},k}
    \end{bmatrix}^T +
    \begin{bmatrix}
    \delta\theta_{\mathrm{w},k+1} & \delta\theta_{\mathrm{r},k+1}
    \end{bmatrix}^T
    \cdot \delta k ~.
\end{align}

Following the dynamics from Equations~(\ref{eq:drivingDynamics_func_discr}) and (\ref{Eq:thermal_dyn_final}), the functions within the MHE dynamics in Equation~(\ref{Eq. nl_discrete}) can be further defined as 
\begin{align}\label{EQ:mhe_model}
\begin{split}
    \Tilde{f}(x_k, w_k) &= f_{\mathrm{DNN}}\big( f_{\mathrm{DD}}(x_k),\ x_k \big) + w_k ~, \\
    g( x_{k}) &= f_{\mathrm{DNN}}( x_k ) ~.        
\end{split}
\end{align}
Furthermore, using these dynamics, the MHE optimization problem including the objective function is defined as: 

\begin{mini}
	{\substack{x_{T-N},\ldots, x_T, \\ w_{T-N},\ldots, w_{T-N}}}
	{
    \underbrace{\frac{1}{2}\|x_{T-N}-\bar{x}_{T-N}\|_{P_0^{-1}}^2}_{\alpha_k(x_k)} + \sum_{k=T-N}^{T-1}
    \frac{1}{2}\|w_k\|_{Q^{-1}}^2
    +
    \frac{1}{2}\|g(x_k)-y_k\|_{R^{-1}}^2} 
	{}
	{\label{Eq. cost_fn}}
	\addConstraint{x_{k+1}}{=\Tilde{f}(x_k, w_k),}{~k=T-N,\ldots,T- 1,}
\end{mini}

Here, \(x = (x_k, \dots, x_{k+N+1})\) represents the optimization variables, while \(y = (y_k, \dots, y_N)\) forms the measurement window.
The weighting matrices \(Q\) and \(R\) are positive definite diagonal matrices corresponding to process and measurement variances; $P_0$ is the weighting matrix of the arrival cost.
MHE operates by optimizing on a fixed horizon of $N$ past measurements $[T-N,T]$, where past information outside the estimation window is not directly included in the optimization (see also Figure~\ref{fig:MHE_concept_intro}). 
The arrival cost, $\alpha_k(x_k)$, addresses this by approximately incorporating information from prior states \([0, T-N-1]\).

This concludes the formulation of the MHE and sets the stage for the implementation into an OCP-structured NLP in the framework \texttt{acados} in the following subsection.

\subsubsection{Implementation in \texttt{acados}}


State estimation is performed by minimizing the variance between system predictions and measurements.
To  solve the MHE optimization problem, the optimal value of the additive process noise $w$ must be determined to yield the most accurate estimate. 
The noise $w$ is thus treated explicitly as an optimization variable.

This leads to the optimization problem to be formulated as an OCP, where the process noise $w$ is considered as the controls input.
Here, the OCP is structured around four sets of variables: states $x$, controls $u$, parameters $p$, and measurements $y$, each of which is defined below.

The primary states to be estimated include the winding and rotor temperatures (\(\theta_{\mathrm{w}}, \theta_{\mathrm{r}}\)), with state evolution governed by the DNN-based thermal model. 
Given the recurrent nature of an RNN, its hidden and cell states $(c_k, h_k)$ are incorporated into the state vector, resulting in an increased state dimension that depends on the number of LSTM units (here 8):
\begin{align}
    x_k = \begin{bmatrix}
        \theta_{\mathrm{w},k} & \theta_{\mathrm{r},k} & c_k & h_k 
    \end{bmatrix}^T ~, \quad \in\mathbb{R}^{18} ~.
\end{align}
With the process noise $w_k$ defined as an additive term, the state evolution as seen in Equation~(\ref{EQ:mhe_model}) can be stated as
\begin{align}
    x_{k+1}=\begin{bmatrix}\theta_{\mathrm{w},k+1}+w_{\theta,\mathrm{w},k} & \theta_{\mathrm{r},k+1}+w_{\theta,\mathrm{r},k} & 
    h_{k+1} & 
    c_{k+1} 
    \end{bmatrix}^T = \Tilde{f}(x_k, w_k), \quad \in\mathbb{R}^{4} ~. 
\end{align}

Further, the control vector can be represented as
\begin{equation}
    u_k = w_k = \begin{bmatrix}
        w_{\theta,\mathrm{w},k} & w_{\theta,\mathrm{r},k} 
    \end{bmatrix} ^T ~, \qquad w_k\sim\mathcal{N}(0,Q(\alpha)) ~, \quad \in\mathbb{R}^{2} ~.
\end{equation}
The process noise follows a Gaussian distribution $\mathcal{N}$  with a variance $Q(\alpha)$, a diagonal matrix derived from the variance  of the states.
This variance matrix also serves as a weighting matrix in the optimization of state estimation accuracy (see Equation~(\ref{Eq. cost_fn})).
The MHE model utilizes past control inputs from the MPC, including \(T_{\mathrm{EM,acc}},\) \(T_{\mathrm{EM,brk}},\) \(T_{\mathrm{fric, brk}}\), along with road inclination $\phi$, and vehicle velocity $v_{\mathrm{veh}}$ as inputs to predict the states.
Thus, the parameter vector is defined as
\begin{align}
    p = \begin{bmatrix}
        T_{\mathrm{EM,acc},k} & T_{\mathrm{EM,brk},k} & T_{\mathrm{fric,brk},k} & \phi & v_{\mathrm{veh},k}
    \end{bmatrix} ^T~, \quad \in\mathbb{R}^{5} ~.
\end{align}

The measurements $y$, as defined in Equation~(\ref{Eq. nl_discrete}), incorporate additive sensor noise to account for real-world inaccuracies such as offset and sensitivity errors. 
These errors are modeled using additive white Gaussian noise \cite{haykin09}, along with a time delay to represent thermal lag in sensor response. 
Consequently, the noisy measurements \((\theta_{\mathrm{w,meas}}, \theta_{\mathrm{r,meas}})\) serve as inputs to the MHE:
\begin{align}
    y &=  \begin{bmatrix}
        \theta_{\mathrm{w,meas}} & \theta_{\mathrm{r,meas}}   
    \end{bmatrix}^T ~.
\end{align}

The term $\alpha_k(x_k)$ from Equation~(\ref{Eq. cost_fn}) in the \texttt{acados} OCP framework encapsulates the information required for the initial node computation and arrival cost. The vectors are defined as
\begin{align}
    x_0 &= \begin{bmatrix}
        \theta_{\mathrm{w},k} & \theta_{\mathrm{r},k} & w_{\theta,\mathrm{w}} & w_{\theta,\mathrm{r}} & \theta_{\mathrm{w},k} & \theta_{\mathrm{w},k}
    \end{bmatrix}^T ~, \notag \\
    \overline{x}_0 &= \begin{bmatrix}
        \theta_{\mathrm{w,meas}} & \theta_{\mathrm{r,meas}} & 0 & 0 & \hat{\theta}_{\mathrm{w},k} & \hat{\theta}_{\mathrm{r},k}
    \end{bmatrix}^T ~.
\end{align} 

Here, $\theta_{\mathrm{w},k}, \theta_{\mathrm{r},k}, w_{\theta,\mathrm{w},k-1} ~, w_{\theta,\mathrm{r},k-1}$ form the vector for the first node and the remaining terms serve as input for the arrival cost. 
In the recursive online simulation, the optimizer continuously refines the state trajectory over a moving horizon.

\texttt{acados} performs simulation in a forward time manner, progressing from timestep $k$ to $(k+N)$, which corresponds to $(T-N)$ to $T$ as illustrated in Figure~\ref{fig:MHE_concept_intro}. 
The estimated state at the final node \((k+N)\) serves as the current timestep estimate $T$, while the state at the first node \(k\) contributes to the arrival cost for the next OCP iteration with a shifted horizon. 

The control variables \(u_k\) are optimized to align predictions with available measurements while accounting for uncertainty.
A key feature of this approach is that the optimizer autonomously determines the optimal noise added to the state, eliminating the necessity for explicit constraints on controls.
Constraints on the physical states $( \underline{\theta}_{\mathrm{w}}, \underline{\theta}_{\mathrm{r}}, \overline{\theta}_{\mathrm{w}}, \overline{\theta}_{\mathrm{r}}) $ remain enforced.
The various parameters settings, weighting matrices, and constraints on the OCP model are detailed in Section~\ref{sec:SimResults} along with the results.

\section{Results and Discussion}
{This section presents the results of the proposed framework, beginning with the MiL simulation used to evaluate the performance and robustness of the DNN-based MHE through fault injection tests.
The results also include the implementation of the MHE on an embedded platform, demonstrating its real-time capability and embedded compatibility.
All findings are critically discussed to highlight both strengths and limitations of the approach.}

\label{sec:SimResults}
\subsection{Model-in-Loop Simulation}

\begin{figure}[H]
	\includegraphics[width=0.95 \textwidth]{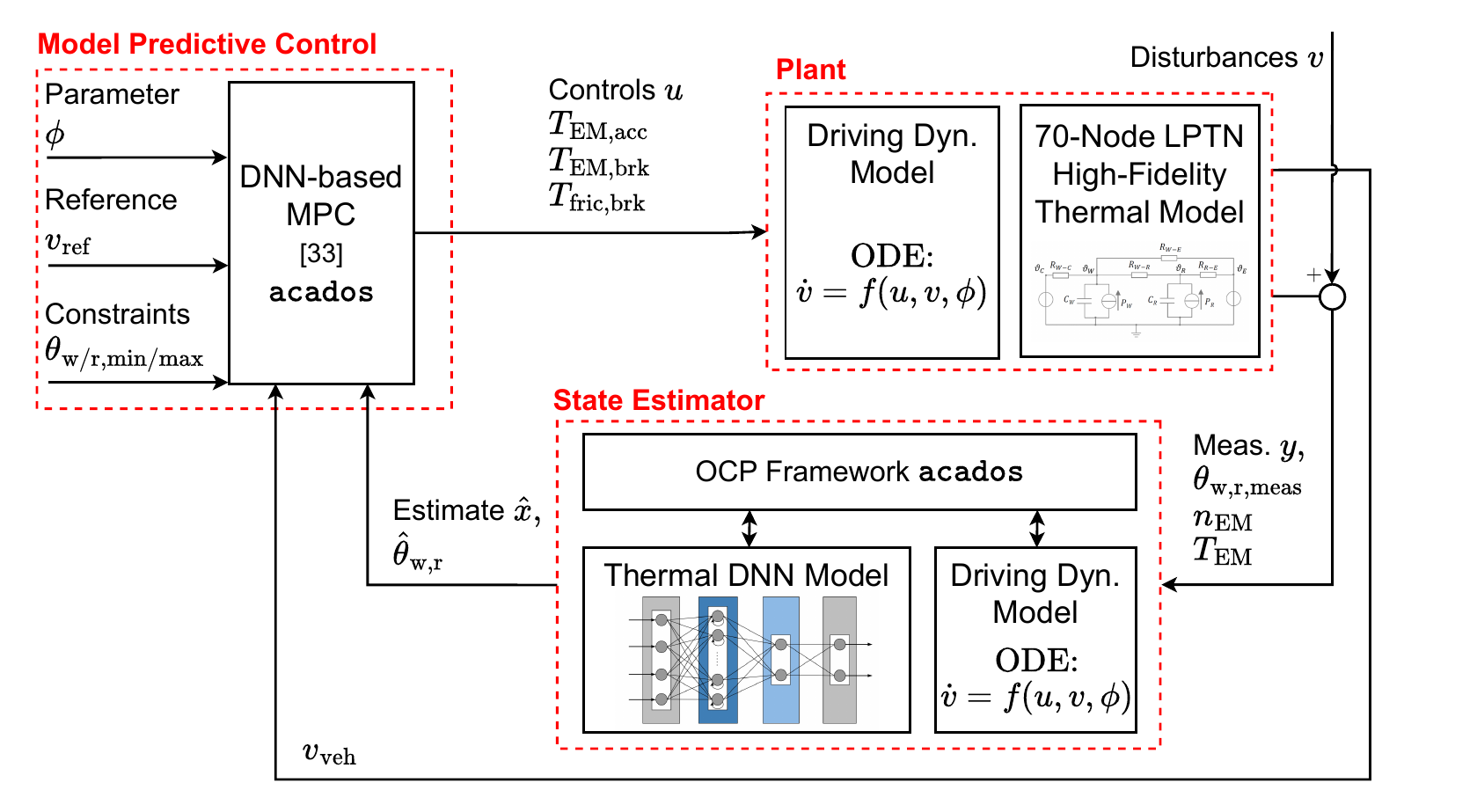}
	\caption[Simulation and model setup for MHE application and validation]{Simulation and model setup for MHE application and validation. DNN-based MPC as presented in Ref.~\cite{WINKLER20238254}.}
	\label{fig:Simloop}
\end{figure}

The MiL simulation is performed over a single lap of the Nürburgring Nordschleife test dataset, with the reference velocity \(v_\mathrm{ref}\) generated offline. 
The MPC tracks \(v_\mathrm{ref}\) while adhering to system constraints and determines the control variables, applied to the vehicle without any control disturbances. 
The resulting torque determines the vehicle velocity through the driving dynamics model, while the corresponding electric machine temperatures are computed using a high-fidelity 70-node LPTN model in the plant.
To simulate realistic measurement conditions, a noise is added to the temperature values obtained from the high-fidelity plant model, yielding imprecise sensor readings.
The overall simulation framework is illustrated in Figure~\ref{fig:Simloop}.

Based on the properties and specifications of commonly used temperature sensors, a negative mean of ${-1}$ $^\circ$C and variance $0.1$ is applied to the added measurement noise thus resulting in \(\nu \sim\mathcal{N}(-1,0.1)\). Additionally, to account for thermal lag, a $1.5$~s delay is incorporated.
The key parameters of the integration of the OCP-structured NLP in the \texttt{acados} framework are summarized in Table~\ref{tab:OCP_param}, using a linear least squares cost function.
The prediction horizon \( T \) and the estimation horizon \( N \) are set to 1.5 s and 15 shooting nodes, respectively, ensuring a balance between prediction accuracy and computational feasibility. 
The weighting matrices \( (Q, R) \) are tuned based on training session data and modeled sensor noise, with a greater emphasis on arrival cost to ensure effective assimilation of past information into the estimation process. 
\vspace{-6pt}

\begin{table}[H]
    \caption{{Tuning parameters for Optimal Control Problem.}}
    \setlength{\tabcolsep}{6.1mm}
    \renewcommand{\arraystretch}{1.2}
    \begin{tabular}{lll}
        \toprule
        \textbf{Symbol} & \textbf{Parameter} & \textbf{Value}\\
        \midrule        $\underline{\theta}_\mathrm{w},\underline{\theta}_\mathrm{r}$ & Minimum winding and rotor temperature & {0} $^\circ$C \\        $\overline{\theta}_\mathrm{w},\overline{\theta}_\mathrm{r}$ & Maximum winding and rotor temperature & {155} $^\circ$C \\
        \(P_0\) & Weighting matrix of the arrival cost & diag(1, 1)\\
        \(Q\) & Weighting matrix of mapped states & \(\text{diag}(0.02, 0.02)\)\\        
        \( R \) & Weighting matrix of controls & \(\text{diag}(0.7, 0.7)\)\\  
        $N$ & Estimation horizon & 15\\        
        $\delta{k}$ & Timestep size & {100} ms \\
        $T$ & Horizon length & {1.5} s \\
        \bottomrule
    \end{tabular}    
    \label{tab:OCP_param}
\end{table}

Sequential Quadratic Programming (SQP) is used to solve the OCP-structured NLP, with a maximum number of SQP iterations of 20. 
The quadratic subproblems are solved using the High-Performance Interior Point Method (HPIPM) framework developed in~\cite{Diehl20} which is interfaced via \texttt{acados}. 
To further reduce computational complexity, the full estimation horizon \( N \) is condensed from 15 to 5 nodes using a partial condensing routine.  

The primary focus of this study is to assess the feasibility of the MHE framework using a DNN-based plant model.
Instead of quantitatively comparing vehicle performance under MPC control, the results are evaluated qualitatively to determine the effectiveness of the DNN-based state estimator in reconstructing temperature states.
Given that only the winding temperature \(\theta_{\mathrm{w}}\) is susceptible to reaching critical thresholds, the evaluation is centred on monitoring this key metric.  

\begin{figure}[H]
\includegraphics[width=0.9\textwidth]{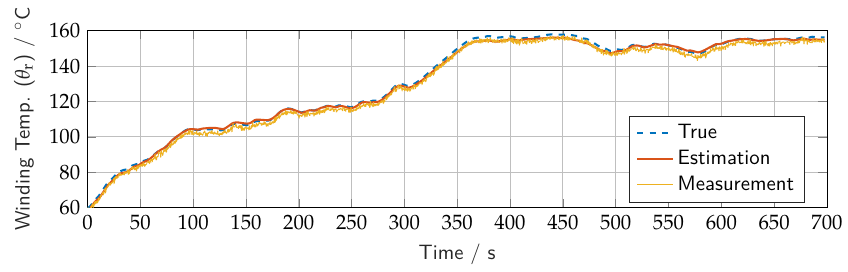}
	\caption{MHE 
  utilizing LSTM neural network compared to true and measured value (incl. noise) for whole drive cycle run.}
	\label{fig:MHE_LSTM_results}
\end{figure}

\begin{figure}[H]
\includegraphics[width=0.9\textwidth]{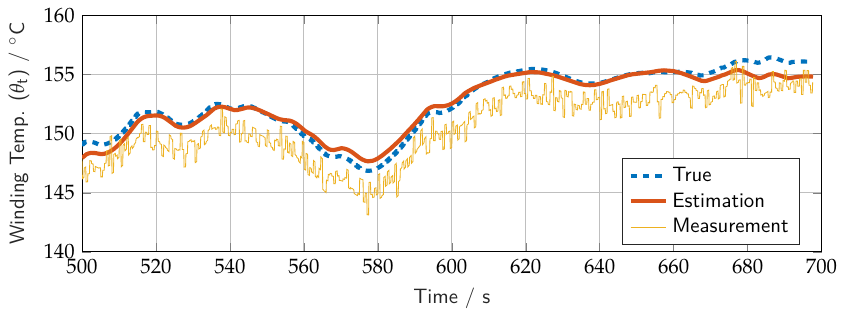}
	\caption{MHE 
  utilizing LSTM neural network compared to true and measured value (incl. noise) for sensible temperature range (zoom).}
	\label{fig:MHE_LSTM_results_zoom}
\end{figure}

Figure~\ref{fig:MHE_LSTM_results} presents the winding temperature estimates generated by the MHE. 
A closer examination in Figure~\ref{fig:MHE_LSTM_results_zoom} focuses on the temperature range of {140} $^\circ$C to {165} $^\circ$C, where excessive heating poses a risk of PMSM damage. 
The estimated values are compared against ideal temperature profiles from the plant model and noisy sensor measurements. 
Despite deviations in raw sensor readings, the DNN-based MHE effectively optimizes state estimates, producing values that closely align with the ideal temperature profile. 

\begin{figure}[H]
\includegraphics[width=0.9\textwidth]{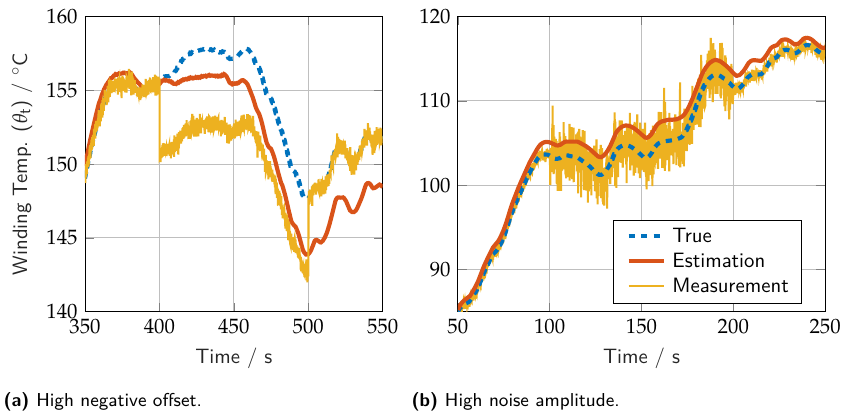}
\caption{Zoom 
 plots of estimator outputs for winding temperature compared to true and measured value (incl. noise) with heavy failure injection for robustness investigation.}
\label{fig:MHE_LSTM_results_noise_zoom}
\end{figure}

To further evaluate the robustness of the DNN-based MHE, an experimental scenario is designed by introducing artificial sensor failures, including a high negative offset and increased noise amplitude. 
As shown in Figure~\ref{fig:MHE_LSTM_results_noise_zoom}, the left plot (a) illustrates the impact of a significant negative offset, while the right plot (b) demonstrates the effect of high noise amplitude.  

Despite these extreme conditions, the MHE maintains stability in its outputs, preventing excessive oscillations or deviations from the true values.
The ability of the DNN-based MHE to provide reliable state estimates under severe sensor disturbances underscores its robustness, reinforcing its suitability for safety-critical applications.

{The results presented focus exclusively on the performance of the estimator, specifically the ability of the DNN-based MHE to reconstruct unmeasured temperature states under realistic measurement conditions and computational constraints.
Metrics such as lap time or energy consumption, which characterize overall vehicle performance, are not evaluated in this study. 
This is a deliberate choice, as the primary objective of this work is to demonstrate a proof of concept for integrating DNNs within a MHE framework and to assess real-time feasibility. 
Accordingly, the vehicle dynamics are used as a validation environment rather than as benchmarks for overall system performance.}

{Future work could involve extending this approach by benchmarking multiple state estimation techniques—both model-based and data-driven—under identical control and vehicle simulation scenarios.
This would enable a comprehensive evaluation not only of estimation accuracy and robustness, but also of their downstream impact on control performance, such as lap time, energy efficiency, and thermal safety margins.}

\subsection{Embedded Integration}
\label{sec:EmbeddedIntegration}

The state estimator undergoes real-time validation following the successful MiL simulations. 
To assess its feasibility for real-world applications, the MHE's real-time performance on an embedded system is evaluated, ensuring computational efficiency and compatibility with production code and control units.

For this purpose, the simulation is deployed on a SCALEXIO 
real-time embedded hardware-in-the-loop (HiL) system from dSPACE.
The processing unit features a {3.8} GHz processor with four cores, three of which are dedicated to model computation.
The MPC and MHE run as separate instances on two cores, while the third core handles the vehicle model, reference trajectory, and system interfacing.
Although this system can be considered as more powerful than traditional processors in automotive engine, vehicle or sensor control units, the MHE problem itself remains computationally demanding due to its high state and control dimensions and horizon length.
The necessary cross-compilation of the libraries is executed based on the \texttt{embedded workflow}, presented by the \texttt{acados}~developers~\cite{acadosEmbeddedWorkflow}.

With a solver timestep of {100} ms, the performance of the solver is evaluated based on the corresponding processor calculation time.
The solver executes a maximum of 20 SQP iterations per timestep, with a peak solver time of {28} ms and an average of {5.7} ms.
Table~\ref{tab:RT_test} shows the relevant parameters and results of the real-time testing.

\begin{table}[H]
 \caption{{Parameter settings and simulation results for embedded real-time testing of deep neural network-based moving horizon estimator.}}
    \label{tab:RT_test}
\setlength{\tabcolsep}{7mm}
    \renewcommand{\arraystretch}{1.2}
    \begin{tabular}{ll}
        \toprule
        \textbf{Parameter} & \textbf{Value} \\
        \midrule
        Timestep & {100} ms \\        
        Horizon length (nodes) & 15, condensed to 5 \\          
        Maximum number of SQP iterations & 20 \\        
        Maximum number of iterations within the QP solver & 100 \\
        Maximum computation time per timestep & {28} ms \\        
        Average computation time per timestep & {5.7} ms \\
        \bottomrule
    \end{tabular}   
\end{table}

{These results indicate that the DNN-based MHE achieves approximately threefold real-time capability, as each control interval permits up to {100} ms for computation, while the solver requires at most {28} ms per step.
This substantial computational margin demonstrates the estimator’s suitability for real-time deployment under the tested configuration.}

{However, it is important to note that the real-time experiments were conducted on a powerful platform equipped with a {3.8} GHz multi-core processor.
The hardware provides significantly greater computational resources than conventional automotive electronic control units (ECUs). 
While this setup is highly effective for prototyping and validating algorithms, it does not reflect the constraints of production-grade automotive hardware.
Consequently, further evaluation on representative low-cost ECUs is essential to assess the estimator’s computational viability and optimize its implementation for series production.}

\section{Conclusions}

This research introduces a novel state estimation framework that integrates DNNs into MHE, replacing conventional physics-based models with data-driven approaches.
This innovation enhances adaptability and computational efficiency, making it suitable for real-time applications.

Using extensive synthetically generated data from a high-fidelity thermal model, a DNN featuring LSTM nodes to enhance its temporal prediction performance is trained.
The MHE is then formulated by integrating the DNN thermal model with one-dimensional driving dynamics in a discrete form, employing forward propagation for the DNN dynamics.
Additionally, the LSTM's hidden and cell states, which capture the long-term dependencies, are incorporated to the MHE's state vector to preserve the DNN's dynamics.
The OCP-structured NLP is then solved using the open-source framework \texttt{acados}.
Through MiL simulations of thermal derating for a PMSM in a BEV, the framework demonstrated accurate estimation of critical temperatures, even under noisy sensor conditions and artificial sensor failures.
Notably, it achieved a three-fold real-time capability on a real-time computer, confirming its feasibility for embedded systems.


{However, several limitations remain. The framework’s performance depends heavily on the quality and coverage of the training data, and its generalization to other systems  remains unverified. 
Additionally, the lack of interpretability in data-driven models may limit adoption in safety-critical applications.
Furthermore, the current embedded implementation is evaluated on high-performance real-time hardware, which may not reflect the constraints of production-grade ECUs. 
Another limitation is the lack of physical performance evaluation metrics as this work primarily serves as a proof of concept for integrating MHE with DNN-based thermal modeling.}

{Future work will expand the evaluation of estimator designs, including comparisons with alternative state estimation methods. Exploring deeper architectures of the DNN may balance estimation quality with computational efficiency, while transfer learning could enhance generalization across systems with minimal retraining.
Incorporating anomaly detection mechanisms may improve fault resilience, and self-learning approaches hold promise for adapting to dynamic system changes without relying on pre-collected data.
Real-time deployment on production-grade ECUs will also advance practical adoption, enabling testing under realistic computational and memory constraints, and further validating the estimator's efficiency and robustness.}


Overall, this research highlights the potential of DNN-based MHE for complex, real-time control applications, particularly in scenarios where accurate mathematical models are difficult to obtain or computationally expensive.
By bridging the gap between model-based and data-driven approaches, this work paves the way for rapidly developed, adaptive, and computationally efficient state estimation frameworks suitable for next-generation, safety-critical systems.
\vspace{6pt}


\authorcontributions{{A.W.}: Writing---Original Draft (lead), Conceptualization, Software, Validation, Visualization, Data Curation; 
{P.S.}: Methodology, Investigation, Software (lead), Writing---Original Draft, Data Curation; 
{K.B.}: Methodology, Writing---Review and Editing; 
{V.S.}: Validation, Writing---Review and Editing, Data Curation; 
{D.G.}: Supervision, Validation, Writing---Review and Editing; 
{J.A.}: Supervision, Project Administration, Funding Acquisition, Writing---Review and Editing. All authors have read and agreed to the published version of the manuscript.}

\funding{The authors disclosed receipt of the following financial support for the research, authorship, and/or publication of this article: The research was performed as part of the Research Group (Forschungsgruppe) FOR 2401 “Optimization based Multiscale Control for Low Temperature Combustion Engines,” which is funded by the German Research Association (Deutsche Forschungsgemeinschaft, DFG).}

\dataavailability{The datasets and scripts presented in this work are publicly available on Zenodo~\cite{Winkler2025MheZenodo}, accessed on 8 July 2025, at 
\url{https://zenodo.org/records/15165902}, including the synthetic training data, training scripts, code for MPC and MHE generation using \texttt{acados}, and the simulation.}


\conflictsofinterest{The authors declare that they have no known competing financial interests or personal relationships that could have appeared to influence the work reported in this paper.}



\begin{thebibliography}{999}

\bibitem[Simon(2006)]{Simon2006}
Simon, D.
\newblock {\em Optimal State Estimation: Kalman, H Infinity, and Nonlinear
  Approaches}; John Wiley \& Sons:  Hoboken, NJ, USA, 
 2006.

\bibitem[Aldrich(1997)]{Aldrich97}
Aldrich, J.
\newblock R.A. Fisher and the making of maximum likelihood 1912-1922.
\newblock {\em Stat. Sci.} {\bf 1997}, {\em 12}, 162--176.
\newblock {\url{https://doi.org/10.1214/ss/1030037906}}.

\bibitem[Janacek(1975)]{Janacek75}
Janacek, G.J.
\newblock Estimation of the minimum mean square error of prediction.
\newblock {\em Biometrika} {\bf 1975}, {\em 62},~175.
\newblock {\url{https://doi.org/10.2307/2334501}}.

\bibitem{ribeiro04}
Ribeiro, M. Isabel, 
\newblock \textit{Kalman and Extended Kalman Filters: Concept, Derivation and Properties}, 
\newblock Technical Report, Instituto de Sistemas e Robótica, Instituto Superior Técnico, Lisbon, Portugal, 2004. 
\newblock \url{https://www.researchgate.net/publication/2888846_Kalman_and_Extended_Kalman_Filters_Concept_Derivation_and_Properties}

\bibitem[Julier and Uhlmann(1997)]{Julier97}
Julier, S.; Uhlmann, J.
\newblock New extension of the Kalman filter to nonlinear systems.
\newblock In Proceedings of the SPIE 3068, Signal Processing, Sensor Fusion, and Target Recognition VI, Orlando, FL, USA, 28 July 1997. 
\newblock {\url{https://doi.org/10.1117/12.280797}}.

\bibitem[Rao et~al.(2001)Rao, Rawlings, and Lee]{Rawlings2001}
Rao, C.V.; Rawlings, J.B.; Lee, J.H.
\newblock Constrained linear state estimation---A moving horizon approach.
\newblock {\em Automatica} {\bf 2001}, {\em 37},~1619--1628.
\newblock {\url{https://doi.org/10.1016/s0005-1098(01)00115-7}}.

\bibitem[Vandersteen et~al.(2013)Vandersteen, Diehl, Aerts, and
  Swevers]{Vandersteen2013}
Vandersteen, J.; Diehl, M.; Aerts, C.; Swevers, J.
\newblock Spacecraft Attitude Estimation and Sensor Calibration Using Moving
  Horizon Estimation.
\newblock {\em J. Guid. Control. Dyn.} {\bf 2013}, {\em
  36},~734--742.
\newblock {\url{https://doi.org/10.2514/1.58805}}.

\bibitem[Bae and and(2017)]{Bae03072017}
Bae, H.;  Oh, J.H.
\newblock Humanoid state estimation using a moving horizon estimator.
\newblock {\em Adv. Robot.} {\bf 2017}, {\em 31},~695--705.
\newblock {\url{https://doi.org/10.1080/01691864.2017.1326317}}.

\bibitem[Baumg{\"a}rtner et~al.(2019)Baumg{\"a}rtner, Zanelli, and
  Diehl]{Baumgaertner2019}
Baumg{\"a}rtner, K.; Zanelli, A.; Diehl, M.
\newblock Zero-Order Moving Horizon Estimation.
\newblock In {Proceedings} 
 of the IEEE Conference on Decision and Control, Nice, France, 11--13 December 
 2019.

\bibitem[Baumgärtner et~al.(2021)Baumgärtner, Frey, Hashemi, and
  Diehl]{Baumgaertner2021}
Baumgärtner, K.; Frey, J.; Hashemi, R.; Diehl, M.
\newblock Zero-order moving horizon estimation for large-scale nonlinear
  processes.
\newblock {\em Comput. Chem. Eng.} {\bf 2021}, {\em
  154},~107433.
\newblock
  {\url{https://doi.org/10.1016/j.compchemeng.2021.107433}}.

\bibitem[Girrbach(2021)]{Girrbach2021}
Girrbach, F.
\newblock Moving Horizon Estimation for Inertial Motion Tracking : Algorithms
  and Industrial Applications.
\newblock Ph.D. Thesis, Albert-Ludwigs-Universit{\"a}t Freiburg, Freiburg, Germany, 
  2021.
\newblock {\url{https://doi.org/10.6094/UNIFR/226677}}.

\bibitem[Rawlings and Allan(2021)]{Rawlings2021}
Rawlings, J.B.; Allan, D.A. Moving Horizon Estimation.
\newblock In {\em Encyclopedia of Systems and Control}; Springer International
  Publishing: Cham, Switzerland, 2021; pp. 1352--1358.
\newblock {\url{https://doi.org/10.1007/978-3-030-44184-5_4}}.


\bibitem[Asch et~al.(2016)Asch, Bocquet, and Nodet]{Asch16}
Asch, M.; Bocquet, M.; Nodet, M.
\newblock \emph{Data Assimilation: Methods, Algorithms, and Applications}.
\newblock Volume~11 of \emph{Fundamentals of Algorithms}.
\newblock Society for Industrial and Applied Mathematics (SIAM), Philadelphia, PA, USA, 2016. 
\newblock ISBN: 978-1-61197-453-9. \url{https://doi.org/10.1137/1.9781611974546}.


\bibitem[Brunton and Kutz(2019)]{brunton19}
Brunton, S.L.; Kutz, J.N.
\newblock {\em Data-Driven Science and Engineering}; Cambridge University
  Press: Cambridge, UK, 2019.
\newblock {\url{https://doi.org/10.1017/9781108380690}}.

\bibitem[Goodfellow et~al.(2016)Goodfellow, Bengio, and
  Courville]{Goodfellow16}
Goodfellow, I.; Bengio, Y.; Courville, A.
\newblock {\em Deep Learning}; MIT Press:  Cambridge, MA, USA, 
 2016.
\newblock \url{http://www.deeplearningbook.org}.

\bibitem[Hewing et~al.(2020)Hewing, Wabersich, Menner, and
  Zeilinger]{hewing2020learning}
Hewing, L.; Wabersich, K.P.; Menner, M.; Zeilinger, M.N.
\newblock Learning-based model predictive control: Toward safe learning in
  control.
\newblock {\em Annu. Rev. Control. Robot. Auton. Syst.} {\bf
  2020}, {\em 3},~269--296.
\newblock {\url{https://doi.org/10.1146/annurev-control-090419-075625}}.

\bibitem[Salzmann et~al.(2024)Salzmann, Arrizabalaga, Andersson, Pavone, and
  Ryll]{salzmann2024learning}
Salzmann, T.; Arrizabalaga, J.; Andersson, J.; Pavone, M.; Ryll, M.
\newblock Learning for {CasADi}: {D}ata-driven models in numerical optimization.
\newblock In {\em Proceedings of the 6th Annual Learning for Dynamics and Control Conference},
  Oxford, UK, 2024; 
\newblock Volume 242 of {\em Proceedings of Machine Learning Research},
  pp. 541--553.

\bibitem[Lahr et~al.(2025)Lahr, Näf, Wabersich, Frey, Siehl, Carron, Diehl,
  and Zeilinger]{lahr2024l4acados}
Lahr, A.; Näf, J.; Wabersich, K.P.; Frey, J.; Siehl, P.; Carron, A.; Diehl,
  M.; Zeilinger, M.N.
\newblock L4acados: Learning-based models for acados, applied to Gaussian
  process-based predictive control. \emph{arXiv}  \textbf{2025}, arXiv:2411.19258.

\bibitem[Lecun et~al.(1998)Lecun, Bottou, Bengio, and Haffner]{lecun1998}
Lecun, Y.; Bottou, L.; Bengio, Y.; Haffner, P.
\newblock Gradient-based learning applied to document recognition.
\newblock {\em Proc.  IEEE} {\bf 1998}, {\em 86},~2278--2324.
\newblock {\url{https://doi.org/10.1109/5.726791}}.

\bibitem[Bishop(2006)]{bishop}
Bishop, C.M.
\newblock {\em Pattern Recognition and Machine Learning (Information Science
  and Statistics)}; Springer: Berlin/Heidelberg, Germany, 2006.

\bibitem[Sarker(2021)]{Sarker21}
Sarker, I.H.
\newblock Deep Learning: A comprehensive overview on techniques, taxonomy, applications and research directions.
\newblock {\em SN Comput. Sci.} {\bf 2021}, \emph{2}, 420.
\newblock {\url{https://doi.org/10.1007/s42979-021-00815-1}}.

\bibitem[Hochreiter and Schmidhuber(1997)]{hochreiter1997}
Hochreiter, S.; Schmidhuber, J.
\newblock Long short-term memory.
\newblock {\em Neural Comput.} {\bf 1997}, \emph{9}, 1735--1780.

\bibitem[Bragantini et~al.(2021)Bragantini, Baroli, Posada-Moreno, and
  Benigni]{Bragantini21}
Bragantini, A.; Baroli, D.; Posada-Moreno, A.F.; Benigni, A.
\newblock Neural-network-based state estimation: The effect of pseudo- measurements.
\newblock In Proceedings of the 2021 {IEEE} 30th International Symposium on Industrial Electronics (ISIE), Kyoto, Japan, 20--23 June 
 { 2021}.
\newblock {\url{https://doi.org/10.1109/isie45552.2021.9576442}}.

\bibitem[Suykens et~al.(1995)Suykens, De~Moor, and Vandewalle]{Suykens95}
Suykens, J.a.K.; De~Moor, B.L.R.; Vandewalle, J.
\newblock Nonlinear system identification using neural state space models,
  applicable to robust control design.
\newblock {\em Int. J. Control} {\bf 1995}, {\em
  62},~129--152.
\newblock {\url{https://doi.org/10.1080/00207179508921536}}.

\bibitem[Pan et~al.(2000)Pan, Sung, and Lee]{pan00}
Pan, Y.; Sung, S.W.; Lee, J.H.
\newblock Nonlinear dynamic trend modeling using feedback neural networks and
  prediction error minimization.
\newblock {\em IFAC Proc. Vol.} {\bf 2000}, {\em 33},~827--832.

\bibitem[Mobeen et~al.(2025)Mobeen, Cristobal, Singoji, Rassas, Izadi, Shayan,
  Yazdanshenas, Sohi, Barnsley, Elliott, et~al.]{mobeen2025neural}
Mobeen, S.; Cristobal, J.; Singoji, S.; Rassas, B.; Izadi, M.; Shayan, Z.;
  Yazdanshenas, A.; Sohi, H.K.; Barnsley, R.; Elliott, L.;  et~al.
\newblock Neural Moving Horizon Estimation: A Systematic Literature Review.
\newblock {\em Electronics} {\bf 2025}, {\em 14},~1954.
\newblock {\url{https://doi.org/10.3390/electronics14101954}}.

\bibitem[Song et~al.(2023)Song, Fang, and Huang]{Song.2023}
Song, R.; Fang, Y.; Huang, H.
\newblock Reliable Estimation of Automotive States Based on Optimized Neural
  Networks and Moving Horizon Estimator.
\newblock {\em IEEE/ASME Trans. Mechatron.} {\bf 2023}, {\em
  28},~3238--3249.
\newblock {\url{https://doi.org/10.1109/TMECH.2023.3262365}}.

\bibitem[Mostafavi et~al.(2022)Mostafavi, Doddi, Kalyanam, and
  Schwartz]{Mostafavi.18112022}
Mostafavi, S.; Doddi, H.; Kalyanam, K.; Schwartz, D.
\newblock Nonlinear Moving Horizon Estimation and Model Predictive Control for
  Buildings with Unknown HVAC Dynamics.
\newblock {\em IFAC-Pap.} {\bf 2022}, {\em 55},~71--76.
  {\url{https://doi.org/10.1016/j.ifacol.2023.01.105}}.

\bibitem[Chen et~al.(2021)Chen, Li, Chen, Ren, and Gao]{Chen.2021}
Chen, Y.; Li, C.; Chen, S.; Ren, H.; Gao, Z.
\newblock A Combined Robust Approach Based on Auto-Regressive Long Short-Term
  Memory Network and Moving Horizon Estimation for State-of-Charge Estimation
  of Lithium-Ion Batteries.
\newblock {\em Int. J. Energy Res.} {\bf 2021}, {\em
  45},~12838--12853.
\newblock {\url{https://doi.org/10.1002/er.6615}}.

\bibitem[Alessandri et~al.(12/9/2008 - 12/11/2008)Alessandri, Baglietto,
  Battistelli, and Zoppoli]{Alessandri.129200812112008}
Alessandri, A.; Baglietto, M.; Battistelli, G.; Zoppoli, R.
\newblock Moving-horizon state estimation for nonlinear systems using neural networks.
\newblock In {Proceedings} 
 of the 2008 47th IEEE Conference on Decision and
  Control, Cancun, Mexico, 
  9--11 December 2008; pp. 2557--2562.
\newblock {\url{https://doi.org/10.1109/CDC.2008.4739462}}.

\bibitem[Norouzi et~al.(2022)Norouzi, Shahpouri, Gordon, Winkler, Nuss, Abel,
  Andert, Shahbakhti, and Koch]{NOROUZI2022CEP}
Norouzi, A.; Shahpouri, S.; Gordon, D.; Winkler, A.; Nuss, E.; Abel, D.;
  Andert, J.; Shahbakhti, M.; Koch, C.R.
\newblock Deep learning based model predictive control for compression ignition
  engines.
\newblock {\em Control Eng. Pract.} {\bf 2022}, {\em 127},~105299.
\newblock
  {\url{https://doi.org/10.1016/j.conengprac.2022.105299}}.

\bibitem[Gordon et~al.(2024)Gordon, Winkler, Bedei, Schaber, Pischinger,
  Andert, and Koch]{Gordon2024Intro}
Gordon, D.C.; Winkler, A.; Bedei, J.; Schaber, P.; Pischinger, S.; Andert, J.;
  Koch, C.R.
\newblock Introducing a Deep Neural Network-Based Model Predictive Control Framework for Rapid Controller Implementation.
\newblock In {Proceedings} 
 of the 2024 American Control Conference (ACC), Toronto, ON, Canada, 10--12 July 
 2024;  pp. 5232--5237.
\newblock {\url{https://doi.org/10.23919/ACC60939.2024.10644830}}.

\bibitem[Winkler et~al.(2023)Winkler, Wang, Norouzi, Gordon, Koch, and
  Andert]{WINKLER20238254}
Winkler, A.; Wang, W.; Norouzi, A.; Gordon, D.; Koch, C.; Andert, J.
\newblock Integrating Recurrent Neural Networks into Model Predictive Control
  for Thermal Torque Derating of Electric Machines.
\newblock {\em IFAC-Pap.} {\bf 2023}, {\em 56},~8254--8259.
  {\url{https://doi.org/10.1016/j.ifacol.2023.10.1010}}.

\bibitem[Engelhardt(2017)]{Engelhardt.2017}
Engelhardt, T.
\newblock {\em Derating-Strategien f{\"u}r elektrisch angetriebene Sportwagen};
  {Springer Fachmedien Wiesbaden}: Wiesbaden,  2017.
\newblock {\url{https://doi.org/10.1007/978-3-658-18207-6}}.

\bibitem[Etzold et~al.(2019)Etzold, Fahrbach, Klein, Scheer, Guse, Klawitter,
  Pischinger, and Andert]{Etzold.2019b}
Etzold, K.; Fahrbach, T.; Klein, S.; Scheer, R.; Guse, D.; Klawitter, M.;
  Pischinger, S.; Andert, J.
\newblock Function Development with an Electric-Machine-in-the-Loop Setup: A Case Study.
\newblock {\em IEEE Trans. Transp. Electrif.} {\bf 2019}, \emph{5}, 1419--1429.
 \newblock {\url{https://doi.org/10.1109/TTE.2019.2952288}}.

\bibitem[Wallscheid and B{\"o}cker(2017)]{Wallscheid.2017}
Wallscheid, O.; B{\"o}cker, J. (Eds.)
\newblock { Derating of Automotive Drive Systems Using Model Predictive Control}. In Proceedings of the 2017 IEEE International Symposium on Predictive Control of Electrical Drives and Power Electronics (PRECEDE), Pilsen, Czech Republic, 4--6 September 2017; 
  IEEE: Piscataway, NJ, USA, 2017.
\newblock {\url{https://doi.org/10.1109/PRECEDE40549.2017}}.

\bibitem[Verschueren et~al.(2021)Verschueren, Frison, Kouzoupis, Frey, van
  Duijkeren, Zanelli, Novoselnik, Albin, Quirynen, and Diehl]{Verschueren2021}
Verschueren, R.; Frison, G.; Kouzoupis, D.; Frey, J.; van Duijkeren, N.;
  Zanelli, A.; Novoselnik, B.; Albin, T.; Quirynen, R.; Diehl, M.
\newblock acados---A modular open-source framework for fast embedded optimal control.
\newblock {\em Math. Program. Comput.} {\bf 2021}, \emph{14}, 147--183.
\newblock {\url{https://doi.org/10.1007/s12532-021-00208-8}}.

\bibitem[Brownlee(2017)]{Brownlee17}
Brownlee, J.
\newblock {\em Long Short-Term Memory Networks With Python}; Machine Learning
  Mastery,  2017.

\bibitem[Winkler et~al.(2021)Winkler, Frey, Fahrbach, Frison, Scheer, Diehl,
  and Andert]{WINKLER2021359}
Winkler, A.; Frey, J.; Fahrbach, T.; Frison, G.; Scheer, R.; Diehl, M.; Andert,
  J.
\newblock Embedded Real-Time Nonlinear Model Predictive Control for the Thermal
  Torque Derating of an Electric Vehicle.
\newblock {\em IFAC-Pap.} {\bf 2021}, {\em 54},~359--364.
  {\url{https://doi.org/10.1016/j.ifacol.2021.08.570}}.

\bibitem[Kingma and Ba(2015)]{kingma2014adam}
Kingma, D.P.; Ba, J.
\newblock Adam: {A} Method for Stochastic Optimization.
\newblock In Proceedings of the 3rd International Conference on Learning
  Representations, {ICLR}, San Diego, CA, USA, 7--9 May 2015; Conference
  Track Proceedings; Bengio, Y.; LeCun, Y., Eds.,  2015 
. 
\newblock {\url{https://doi.org/10.48550/arXiv.1412.6980}}.

\bibitem[Rawlings et~al.(2017)Rawlings, Mayne, and Diehl]{Rawlings17}
Rawlings, J.; Mayne, D.; Diehl, M.
\newblock {\em Model Predictive Control: Theory, Computation, and Design}; Nob
  Hill Publishing: 
   Madison, WI, USA, 2017.
   

\bibitem[Haykin and Moher(2009)]{haykin09}
Haykin, S.; Moher, M.
\newblock {\em Communication Systems}, 5th ed.; Wiley:  Hoboken, NJ, USA, 
 2009.

\bibitem[Frison and Diehl(2020)]{Diehl20}
Frison, G.; Diehl, M.
\newblock HPIPM: A high-performance quadratic programming framework for model
  predictive control.
\newblock {\em IFAC-Pap.} {\bf 2020}, {\em 53},~6563--6569.
\newblock {\url{https://doi.org/10.1016/j.ifacol.2020.12.073}}.

\bibitem[Frey et~al.()Frey, Hänggi, Winkler, and
  Diehl]{acadosEmbeddedWorkflow}
Frey, J.; Hänggi, S.; Winkler, A.; Diehl, M.
\newblock Embedded Workflow---acados Documentation.
\newblock Available online:
  \url{https://docs.acados.org/embedded_workflow/index.html} (accessed on 5 March 2025).

\bibitem[Winkler(2025)]{Winkler2025MheZenodo}
Winkler, A.
\newblock Deep Neural Network Based Moving Horizon Estimation: Data, Models, Scripts (feat. acados).  2025.
\newblock Zenodo Repository. Available online: \url{https://doi.org/10.5281/zenodo.15056783} accessed on 10 June 2025). 

\end{thebibliography}
\reftitle{References}

\begin{adjustwidth}{-\extralength}{0cm}

\PublishersNote{}

\end{adjustwidth}

\end{document}